# Changes in the coercivity fields of magnetoresistance hysteresis loops under the influence of a spin-polarized current

E. Yu. Beliayev[1], I. G. Mirzoiev[1], V. V. Andrievskii[1], A. V. Terekhov[1], Yu. A. Kolesnichenko[1], V. A. Horielyi[1], I. A. Chichibaba[2]

[1]B. Verkin Institute for Low Temperature Physics and Engineering

of the National Academy of Sciences of Ukraine, Kharkiv 61103, Ukraine.

[2]National Technical University «Kharkiv Polytechnic Institute», Kharkiv 61100, Ukraine

Abstract

Using the example of a pressed sample consisting of chromium dioxide nanoparticles coated with insulating shells, we study the relationship between the electronic transport system and magnetic subsystem in granular spin-polarized metals. It is shown that the spin-polarized tunneling transport current can affect the coercivity fields of the percolation cluster formed in the sample with decreasing temperature.

Chromium dioxide ($CrO_2$) is the main substance used in magnetic information recording media (magnetic tapes, disk drives, MRAM memory). Increasing its coercive force is essential since it directly affects the recording density and storage time. Among the factors influencing the magnitude of the coercive force, the main ones are temperature, the size of ferromagnetic particles (the value of $H_c$ increases with decreasing particle's size and reaches its maximum for single-domain particles [1, 2]), and their chemical composition (for example, the effect of iron impurities on the value $H_c$ was studied in [3 – 6].

It should be noted that $CrO_2$ possesses not only a metallic type of conductivity but it is also a ferromagnet with Curie temperature $T_C \approx 390$ K and a very nontrivial band structure [7]. Its *s*-electrons are highly localized, providing chemical bonds between chromium and oxygen atoms, and do not participate in conductivity. Thus, chromium dioxide is a true *d*-metal. Furthermore, at zero temperature, the density of states at the Fermi level goes to zero for those of its *d*-electrons which are in the "spin-down" state. Thus, at low enough temperatures, metallic conduction is carried out mainly by spin-polarized *d*-electrons in the "spin up" state (with spin direction collinear to the magnetization vector). For this reason, chromium dioxide belongs to a very small special group of ferromagnets, called "half-metals". The spin polarization of conduction electrons increases with decreasing temperature, and in the limit $T \to 0$, only half of its *d*-electrons being in the "spin up" state participate in conductivity.

At low temperatures, the primary mechanism for conductivity in the studied chromium dioxide composite samples is spin-dependent electron tunnelling between $CrO_2$ grains. At the same time, as will be shown below, the conductivity is also affected by the domain walls' movement inside individual granules, and in some cases the ferroelectric properties of $Cr_2O_3$ intergranular dielectric barriers also matters.

Of course, such unusual electronic transport properties of this fairly widespread compound attract the attention of researchers from both applied and fundamental points of view in connection with the development of spin-dependent electronics - spintronics. Spin-polarized electron transport manifests itself most clearly in the effect of giant tunneling magnetoresistance.

This work is devoted to the features in hysteretic behavior of magnetoresistance of pressed powder composed of $CrO_2$ nanoparticles. To minimize the effect associated with magnetoresistance anisotropy [8], most of the measurements were carried out on samples made of rounded chromium dioxide nanoparticles [9]. However, for acicular nanoparticles, all the results obtained in this work are also valid. The test samples consisted of pressed rounded $CrO_2$ particles with a diameter of 120 nm. The particles' surface was modified during preparation by surface reduction of chromium oxide-IV to the state of oxyhydroxide β-CrOOH, which is an antiferromagnetic insulator [10]. The thickness of β-CrOOH insulating surface layer was 3.6 nm. From tablets obtained by uniaxial cold compression under 5 MPa, a rectangular sample with dimensions $1.5 \times 2 \times 10$ mm$^3$ was cut. Current contacts were made by vacuum deposition of silver on the opposite ends of the parallelepiped, and the potential contacts were made in the form of narrow stripes on one of the side faces. In fact, the sample under study was a random network of tunnel contacts between magnetic nanoparticles. And the hysteretic behavior of the sample's magnetization led to the hysteretic dependences for its magnetoresistance (**Fig. 1**).

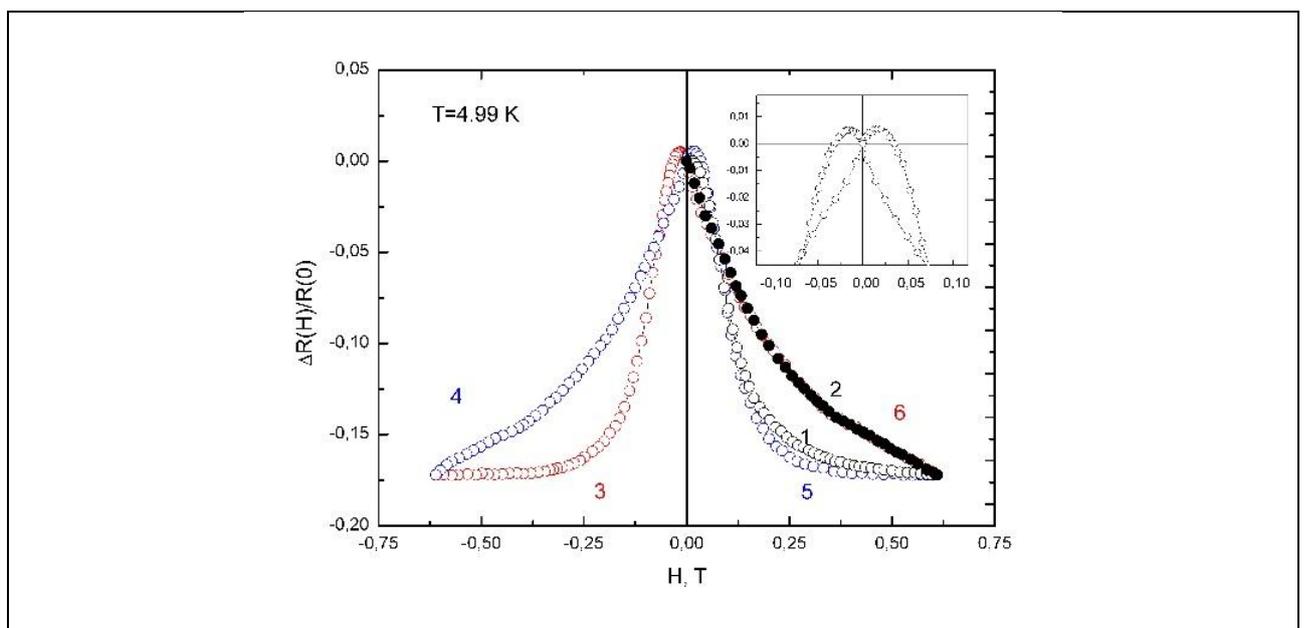

**Fig. 1**. Hysteretic dependences of magnetoresistance for pressed powder sample studied at $T = 4.99$ K. Curves № 1 and № 2 are virgin magnetization curves. The inset shows an enlarged view of the magnetoresistance behavior in low fields.

In the simplest model, a completely disordered state of magnetic moments of ferromagnetic metal granules should correspond to the magnetoresistance maxima observed on the magnetoresistance hysteresis loops at the applied field corresponding to the coercive field $H_c$ (**Fig. 1**). At high enough (room) temperatures, this is observed experimentally. However, with decreasing temperature, the magnetoresistance maxima diverge so that the field of the maxima ($H_p$) exceeds the value of the coercive force ($H_c$) (**Fig. 2a**), and then (for $T<50$ K) the value of $H_p$ begins to decrease sharply (**Fig. 2b**).

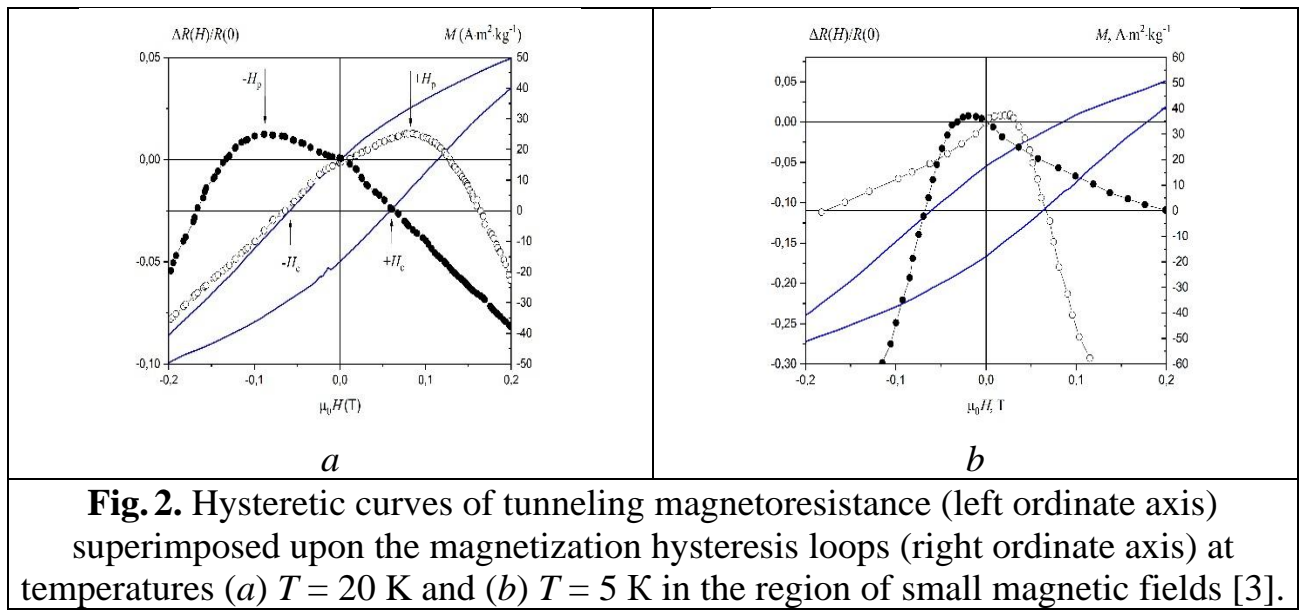

**Fig. 2.** Hysteretic curves of tunneling magnetoresistance (left ordinate axis) superimposed upon the magnetization hysteresis loops (right ordinate axis) at temperatures (*a*) $T = 20$ K and (*b*) $T = 5$ K in the region of small magnetic fields [3].

The temperature dependences for the fields of the magnetoresistance maxima ($H_p$) and the coercive force ($H_c$) for three pressed powder samples made of isolated ferromagnetic $CrO_2$ nanoparticles, differing in shape (acicular or globular), material ($Cr_2O_3$ or $\beta$-CrOOH) of the covering dielectric shells, and thickness of the dielectric coating (a total of 12 similar samples were studied) are shown in Fig. 3. There are significant discrepancies between the fields of the coercivity of the magnetic subsystem $H_c$ of the sample and the magnitudes of the fields of the maxima of the magnetoresistance $H_p$. Their ratio changes with temperature.

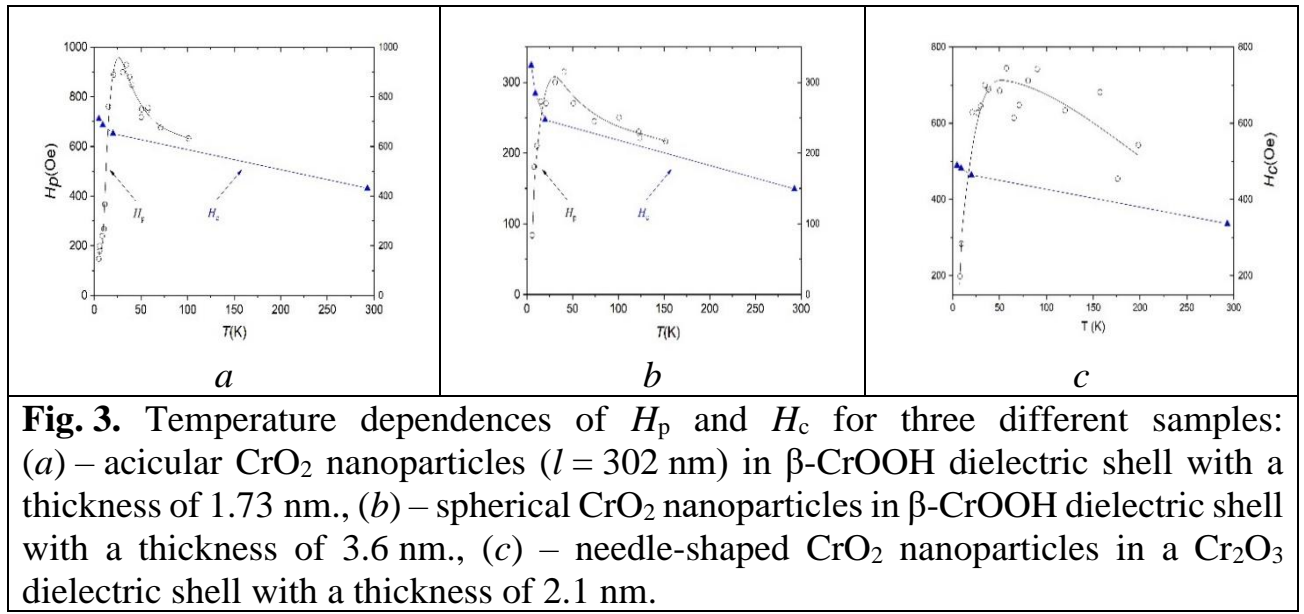

**Fig. 3.** Temperature dependences of $H_p$ and $H_c$ for three different samples: (*a*) – acicular $CrO_2$ nanoparticles ($l = 302$ nm) in β-CrOOH dielectric shell with a thickness of 1.73 nm., (*b*) – spherical $CrO_2$ nanoparticles in β-CrOOH dielectric shell with a thickness of 3.6 nm., (*c*) – needle-shaped $CrO_2$ nanoparticles in a $Cr_2O_3$ dielectric shell with a thickness of 2.1 nm.

Thus, regardless of the shape of the $CrO_2$ nanoparticles used and regardless of the material and thickness of their surface covering:

1. Above $T = 100$ K, the magnitude of the field of magnetoresistance maxima $H_p$ begins substantially approach the magnitude of the coercive force $H_c$ [11].
2. In the intermediate temperature range with decreasing temperature, both values $H_c$ and $H_p$ grow, but the value of the field $H_p(T)$ grows faster than $H_c(T)$. (**Fig. 3**, **Fig. 2a**) It should be noted that this phenomenon is insufficiently covered in the available literature and raises the question of its causes.
3. At the lowest temperatures ($T < 50$ K), the magnitudes of the fields of magnetoresistance maxima $H_p(T)$ begin to fall abruptly; as a result, being significantly lower than the average values of the coercive force $H_c(T)$ measured by the magnetometer (**Fig. 3**, **Fig. 2b**).

To explain the above features in the behavior of the fields of positive magnetoresistance maxima $H_p(T)$, it is necessary to note that the value of $H_p$, is not the averaged magnetic coercivity field of the entire composite sample ($H_c$) but is the coercivity field for its current-conducting part, – percolation cluster.

At high enough temperatures (including above the Curie temperature, which for our samples is $T_c \approx 390$ K), tunnelling barriers between $CrO_2$ nanoparticles can be easily overcome by electrons, and the electric current is uniformly distributed over the sample. Magnetoresistance for these high temperatures ($T > 150 \div 200$ K) does not exceed one percent, and the coercivity field of the magnetic subsystem of the sample, $H_c$, (for $T < T_c$) and the coercivity field for its conducting part, $H_p$, practically coincide [11].

It is known that the tunneling current exponentially depends on both the distance between particles and the height of potential barriers. Therefore, with decreasing temperature, when thermally assisted processes become rare, the granular system becomes a conducting system with an exponential scatter of parameters, and predominant channels of percolation - a percolation cluster - are formed in it.

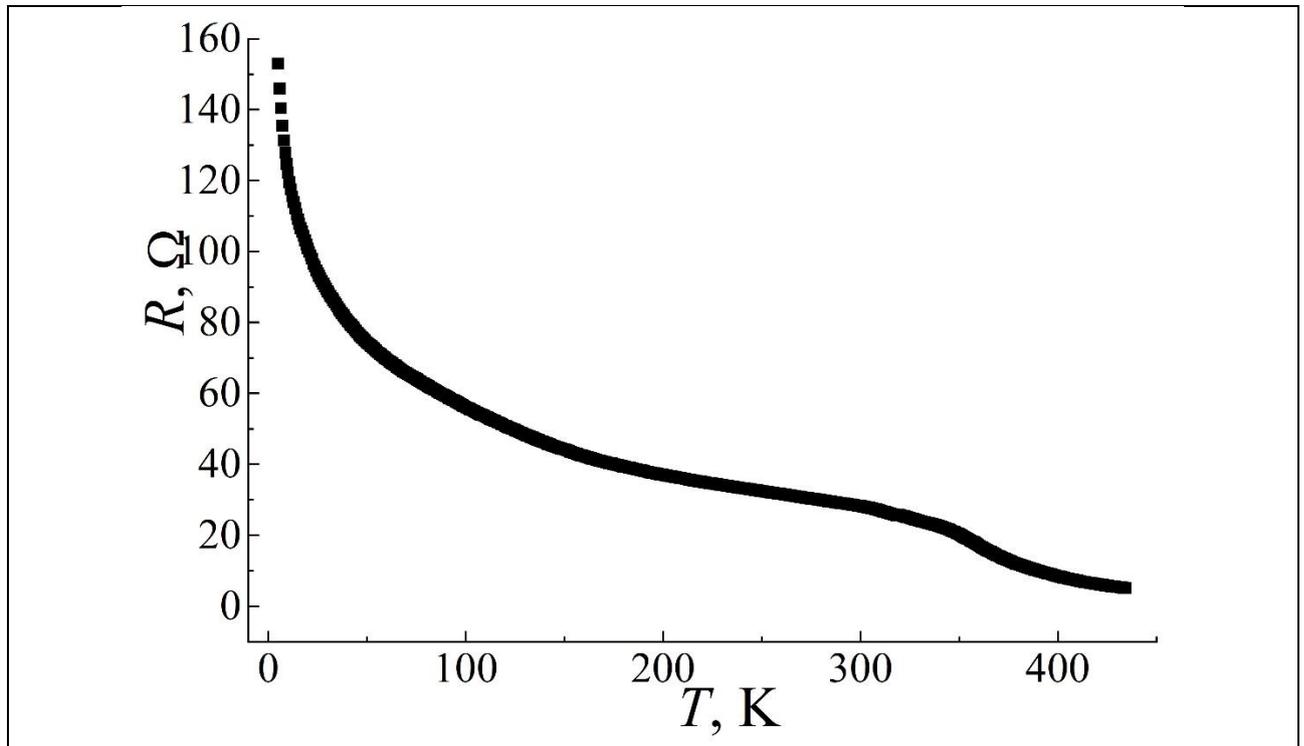

**Fig.4**. Temperature dependence of resistance for the sample under study, consisting of rounded $CrO_2$ nanoparticles, 120 nm in diameter, isolated from each other by β-CrOOH dielectric shells with thickness of 3.6 nm.

In general, the temperature behavior of the resistance for the sample with spherical nanoparticles (see **Fig. 4**) is complex and cannot be described by any single mechanism. Identification of functional contributions of various conduction mechanisms: variable range hopping, nearest neighbor hopping, magnetic scattering and so on (so called, functional deconvolution of the temperature dependence of resistance) [12] is a difficult task and deserves consideration in a separate publication. Here, we just note that $R(T)$ behavior in the high-temperature region is affected by the proximity to FM - PM transition (Curie temperature $T_c = 390$ K), while the low-temperature behavior of the tunneling resistance in a half-metal composite should take into account the growing value of spin polarization in chromium dioxide with decreasing temperature.

Due to the increase in spin polarization ($P$) with decreasing temperature, the tunneling of electrons between isolated $CrO_2$ grains requires not only the presence of states close in energy on both sides of the tunnel barrier but also the possibility of preserving the spin state for the tunneling $d$-electrons. However, the "spin down" states (opposite to the magnetization vector) in $CrO_2$ are localized at the Fermi level with

≈ 1.7 eV dielectric gap in their energy spectrum. In this case, the effect of giant tunneling magnetoresistance arises in the system of electrically isolated ferromagnetic half-metal grains, and the conducting cluster mainly includes links highly ordered in magnetic relation. Such magnetically ordered links are difficult to see in the case of rounded particles (**Fig. 5a**). However, they are clearly visible under an electron microscope in the case of needle-shaped $CrO_2$ nanocrystals, which are oriented in magnetically ordered loops (**Fig. 5b**).

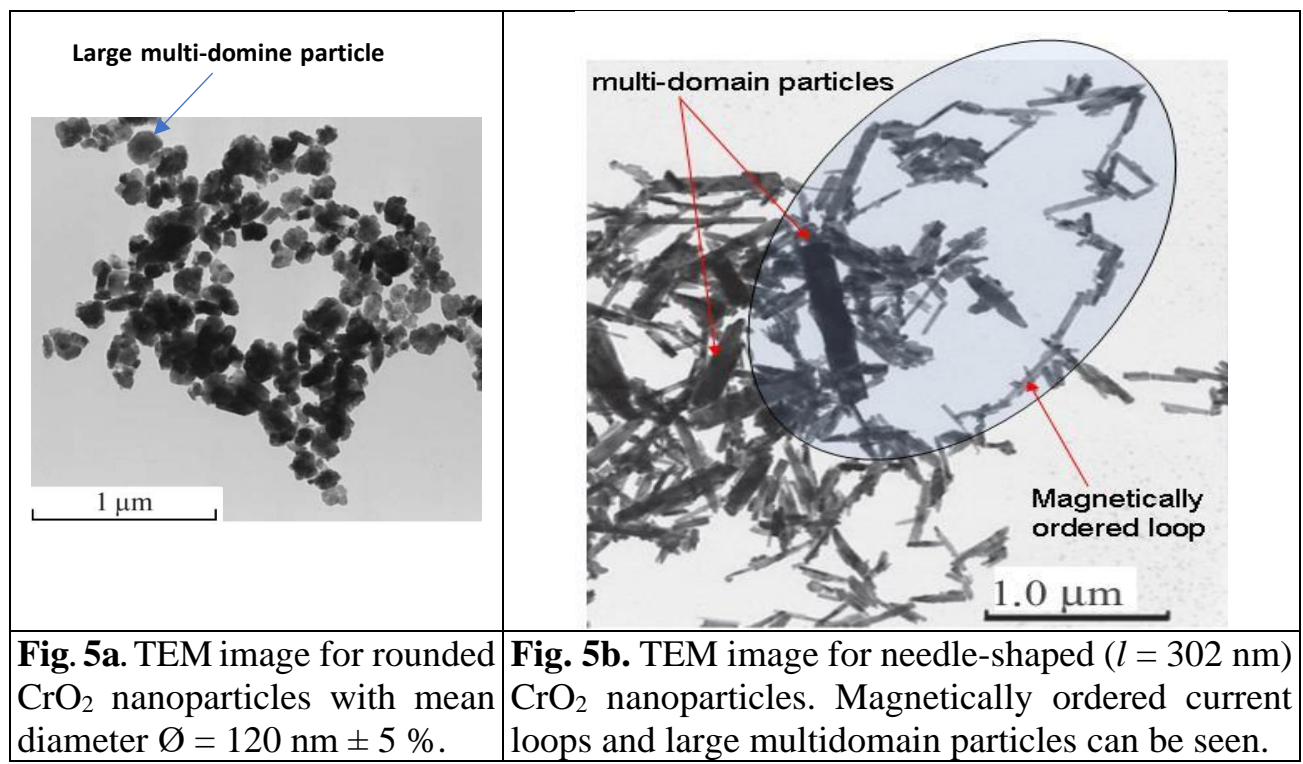

| **Fig. 5a.** TEM image for rounded $CrO_2$ nanoparticles with mean diameter Ø = 120 nm ± 5 %. | **Fig. 5b.** TEM image for needle-shaped ($l$ = 302 nm) $CrO_2$ nanoparticles. Magnetically ordered current loops and large multidomain particles can be seen. |
|---|---|

Obviously, such magnetically ordered regions do not disappear after uniaxial compression during the samples' formation, and they serve as channels for the predominant flow of the spin-polarized current. It is also evident that these highly magnetically ordered percolating current paths have increased coercivity, which is one of the essential factors leading to an excess in the coercivity field $H_p$ for the conducting cluster, compared to the average coercivity field of the magnetic subsystem the whole sample $H_c$.

The behavior of the coercivity fields of the percolation cluster $H_p$ with a further decrease in temperature was considered by us earlier in works [4, 10, 13]. Here we only briefly note that for $T \to 0$, the percolation cluster begins to deplete. More and more dead-end branches appear in it, and the current flow is localized, up to a practically one-dimensional path [14]. Of course, Mott's law cannot be fulfilled in this conduction regime. Furthermore, the growing coefficient of spin polarization of conduction electrons begins to influence the temperature dependence of the resistance. In this case, large multi-domain particles located in the critical links of the current paths acquire special significance. Such big particles, on the one hand, favor tunneling, making a decisive contribution to the conductivity, and on the other hand, being multi-domain by nature, they have a reduced coercivity since the process of their magnetization reversal is not associated with a general changing the direction of their magnetic moment with respect to the axes of magneto-crystalline anisotropy, as is the case for single-domain particles but it is conditioned by the process of the gradual movement of the domain wall, leading to the growth of one domain at the expense of the neighboring one.

It is known that the properties of granular systems significantly depend on the value of the measuring current flowing through the sample [15]. In the above picture, the role of the spin-polarized current remains unclear. Can the flow of a spin-polarized current through a sample affect the coercivity of a percolation cluster? In this regard, we have carried out the measurements and analysis of magnetoresistive hysteresis loops taken in a wide range of temperatures and at different values of the measuring current flowing through the samples.

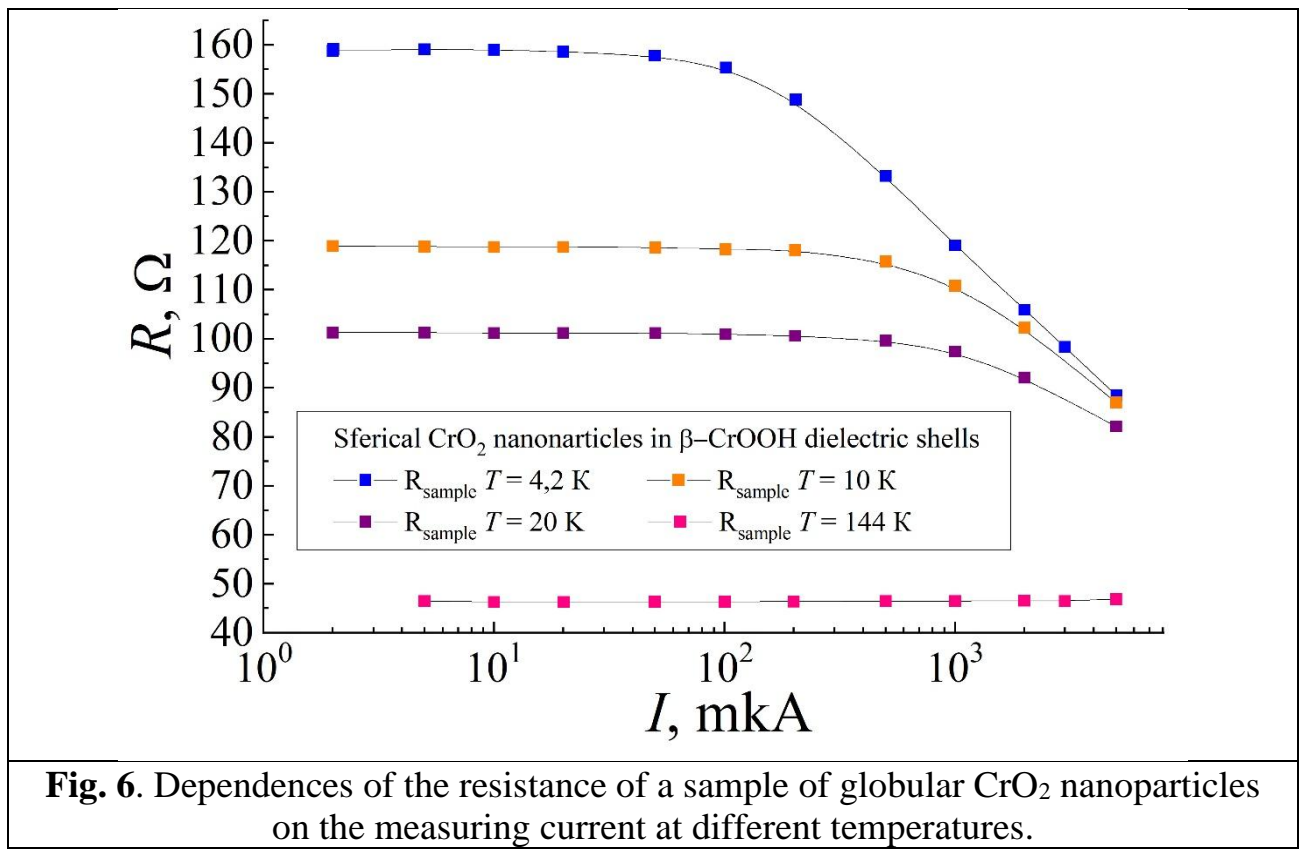

**Fig. 6**. Dependences of the resistance of a sample of globular $CrO_2$ nanoparticles on the measuring current at different temperatures.

According to the results of the analysis of the magnetoresistance hysteresis loops for pressed samples of globular $CrO_2$ nanoparticles in β-CrOOH dielectric shells, it can be concluded that in the intermediate temperature range (50 K < $T$ < 200 K), the excess values of coercivity $H_p$ in comparison with the values of $H_c$ is exclusively due to the initial increased magnetic ordering of the current paths in the branches of the conducting system (percolation cluster). This conclusion is based on:

1. sufficiently uniform current distribution over the sample, as evidenced by the linearity of the *I - V* characteristic measured at $T = 144$ K (**Fig. 6**).
2. the overall low level of magnetoresistance, which, on the one hand, does not allow the position of the magnetoresistance maxima to be estimated with sufficient accuracy (they become very diffuse), and, on the other hand, indicates a general low spin polarization in the samples in this temperature range. It is known that the magnitude of the tunneling magnetoresistance is related to the spin polarization coefficient

$$P = \frac{\rho^\uparrow(\varepsilon_F) - \rho^\downarrow(\varepsilon_F)}{\rho^\uparrow(\varepsilon_F) + \rho^\downarrow(\varepsilon_F)}, \qquad (1)$$

where $\rho^{\uparrow\downarrow}(\varepsilon_F)$ - the density of states for electrons with the Fermi energy $\varepsilon_F$ for the chosen spin direction, by the Julière formula [16]

$$MR = \Delta R(H)/R(0) = P^2/(1 + P^2). \qquad (2)$$

3. the energy of the spin-flip (in fact, the Zeeman splitting energy) is a very small value if compared to the value of the exchange interaction between magnetic ions.

The values of the negative tunneling magnetoresistance, obtained in the maximum magnetic fields available to us ($H = 15$ kOe) at temperatures $T = 4.2$ K, 10 K, 20 K, and 144 K, and the corresponding approximate values of the spin polarization for the sample studied, calculated according to the Julière's formula (2), are given in Table 1.

Table 1.

| $T$, K | $MR = \Delta R(H)/R(0)$, % | $P \approx \sqrt{MR/(1-MR)}$, % |
|---|---|---|
| 144 | 2 | 13 |
| 20 | 17 | 45 |
| 10 | 21 | 52 |
| 4,2 | 36 | 75 |

It should be noted that, of course, the values of spin polarization determined in this way are underestimated for the reasons described in detail in ref. [7], however, they fully demonstrate a tendency for the spin polarization to increase with lowering the temperature. At high temperatures ($T > 50$ K), the fraction of spin-polarized electrons is insignificant, so the effect of the current can be reduced only to Joule heating of the sample, which should lead to a decrease in the coercivity of the magnetoresistance hysteresis loops. However, due to the fairly uniform distribution of the measuring current over the sample, this effect is also insignificant, and within the experimental error (~ 15%), cannot be seen. In this regard, it can be concluded that the increased coercivity fields for magnetoresistance hysteresis loops are solely due to the initial magnetic ordering of $CrO_2$ particles inherent to any ferromagnetic powders (**Fig.5b**).

For these reasons, the effects associated with the influence of the spin-polarized current should be expected only at sufficiently low temperatures, when the percolation cluster is depleted and the level of spin polarization of the tunnel current flowing through the sample increases.

**Fig. 7** shows the dependence of the coercivity fields of the conducting system of the pressed powder sample $H_p$, consisting of rounded half-metal nanoparticles of chromium dioxide $CrO_2$, 120 nm in diameter, isolated by β-CrOOH dielectric surface shells, with thickness 3.6 nm. The measurements were carried out at low temperatures ($T$ = 20, 10, 7, and 4.2 K) and at various values of the measuring current in the range 1 ÷ 5000 μA.

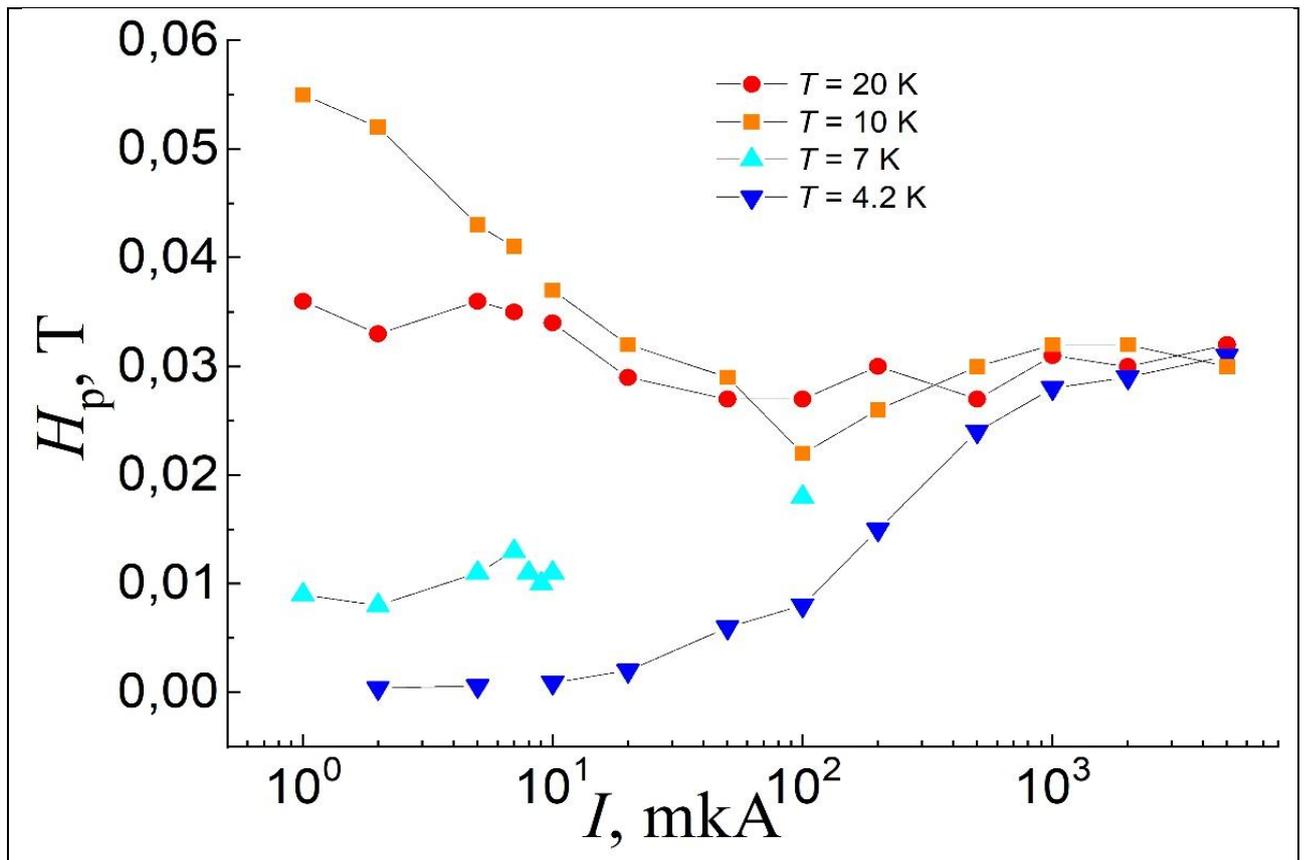

**Fig. 7.** Dependence of the coercivity fields of magnetoresistance hysteresis loops on the measurement current at $T$ = 20, 10, 7, and 4.2 K.

The dependence of the coercivity fields on the measuring current is complex.

At $T = 20$ K, the measuring current is still uniformly distributed over the sample. The $I$-$V$ characteristic of the sample (**Fig. 6**) remains sufficiently linear, and the dependence of the coercivity field $H_p$ on the measuring current is weakly expressed.

With a decrease in temperature to $T = 10$ K, the coercivity of the conducting system increases. However, due to the depletion of the percolation cluster, its critical links begin to overheat (the $I - V$ characteristics shown in **Fig. 6** become nonlinear for $I > 100$ μA), and this overheating, in the limit, leads to a decrease in the field $H_p$ to the level corresponding to $T = 20$ K at $I \approx 100$ μA. With further increase in current the tunnel barriers between the percolation cluster dead branches are suppressed by rising voltages. This leads to the increase in the percolation cluster power (the number of current-carrying channels), and the influence of measuring current on the coercivity fields $H_p$ weakens. Therefore, the curve $H_p(I)$ at $T = 10$ K with a further increase in the measuring current flowing through the conducting cluster (and, accordingly, with an increase in voltage), reduces its slope and merges with the curve for $T = 20$ K.

At $T = 7$ K, the coercivity fields of the magnetoresistance hysteresis loops drop sharply, which can be explained by the predominant influence of large multidomain particles connecting the percolation chains in the conditions of increasing percolation conductivity. According to [10], it is these particles that lead to a decrease in the $H_p$ fields at low temperatures.

At $T = 4.2$ K, the corresponding $H_p$ fields become even smaller.

Attention is drawn to the fact that the current dependence of the field $H_p$ changes its sign at low temperatures. This occurs despite the increasing effect of overheating and the increasing nonlinearity of the $I$ - $V$ characteristics at the lowest temperatures. The coercivity of the percolation cluster increases with increasing current (**Fig. 7**).

This experimental fact suggests that the reason for such an increase in coercivity is the increased spin polarization of the measuring current flowing through the sample.

By itself, even a sufficiently strong spin-polarized current should not affect the magnetic properties of conducting magnetically ordered chains in a percolation cluster. The spin-flip energy is too small compared to the magnetic anisotropy energy of the conducting system. This influence can manifest itself only in critical areas - weak links, through which the main part of current flows. Apparently, such regions are large multidomain particles with reduced coercivity. Note that just by the influence of such particles, according to work [2], the effect of a decrease in the coercivity $H_p$ with lowering temperature can be explained.

It is known that the domain structure of ferromagnets is extremely susceptible to an external magnetic field and other external influences [17]. Including the flow of electric current [18] [19].

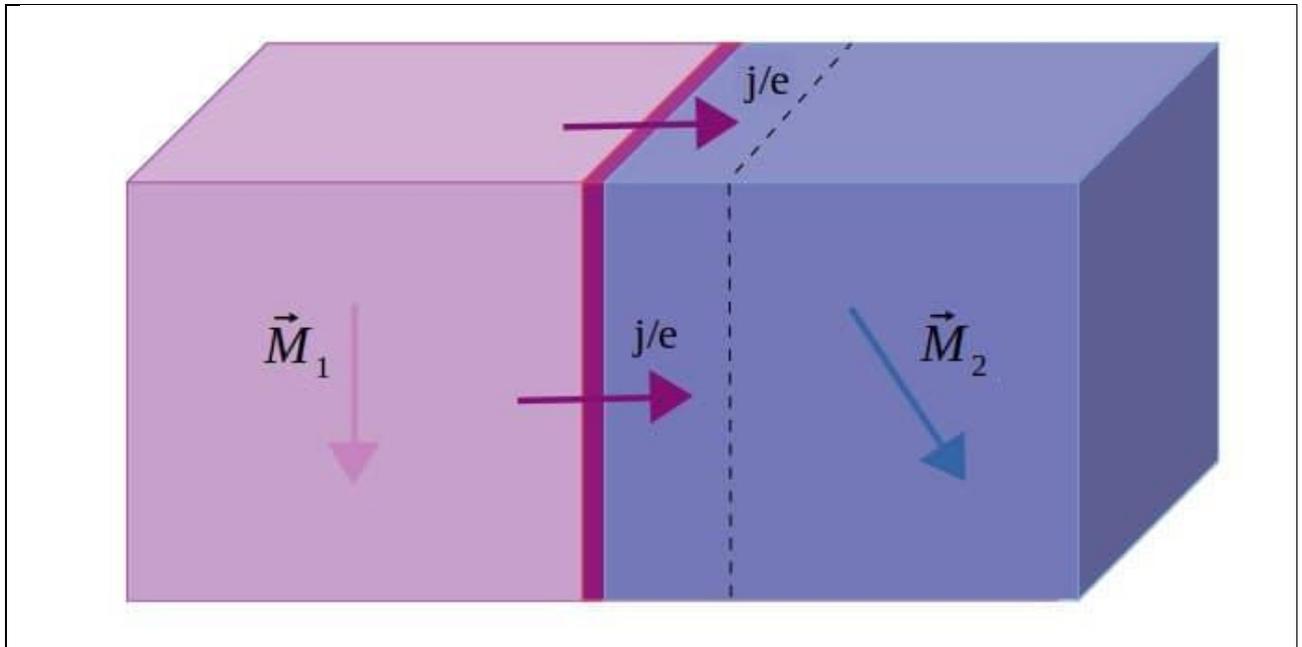

**Fig. 8.** Scheme of spin injection and spin accumulation in a two-domain particle. Spin-polarized current leads to a displacement of the domain wall and increases the coercivity of the conducting system in terms of a response to an external magnetic field.

The paper [20] considers in detail the influence of the spin-polarized current on the motion of the domain wall. **Fig. 8** shows how spin-polarized electrons are injected into an adjacent domain. At sufficiently strong currents, the amount of spin accumulation on the other side of the domain wall increases, which can cause a shift of the domain wall relative to its equilibrium position. According to work [21], the mass of a topological quasiparticle corresponding to a domain wall in a ferromagnetic nanowire $Ni_{81}Fe_{19}$ is $6.6 \times 10^{-23}$ kg. In this case, the current flow with a density $10^{10}$ A×m$^{-2}$ caused a displacement of the domain wall relative to its equilibrium position by 10 μm. In the powders studied in this paper, most of the particles had a rounded shape with a diameter 120 nm and were single-domain. (According to electron microscopic studies, up to 5% of particles showed deviations in size.) The current density for the value of the measuring current 100 μA in most of our experiments, in the limiting case, while flowing through, for example, a two-domain particle with dimensions 120×240 nm$^2$ could be up to $0.3 \times 10^{10}$ A×m$^{-2}$. It is obvious that a spin-polarized current with such a high density can not only shift the domain wall, but also completely rearrange the neighboring domain, turning a rather large chromium dioxide particle into a single-domain one.

Thus, the mechanism for the increase in the coercivity fields under the influence of the spin-polarized current appears to be as follows:

The motion of domain walls in multidomain particles under the influence of weak magnetic fields in equilibrium conditions (in the absence of spin-polarized current) is characterized by high mobility, which corresponds to a low coercivity of magnetoresistance hysteresis loops. With an increase in the magnitude of the spin-polarized current, the growing deviation of domain walls from the equilibrium position leads to a decrease in their response to the influences from the side of an external magnetic field, which corresponds to an increase in the coercivity of the percolation cluster with increasing current.

It should also be noted that all the obtained dependences $H_p(I, T)$ shown in **Fig. 7**, in the limit of strong currents (and, accordingly, high applied voltages leading to a decrease in tunnel barriers and a growth of the percolation cluster) converge at a single intersection point $H_p \approx 300$ Oersted, which approximately corresponds to the value of the average coercive force $H_c$ obtained from magnetometry measurements for the sample studied, consisting of rounded $CrO_2$ nanoparticles (see **Fig. 3b**).

Thus, we have shown that:

1. The excess of the magnitude of the coercivity fields $H_p$ of the magnetoresistance hysteresis loops in comparison with the magnitude of the coercive force $H_c$, measured by magnetometry methods, in the intermediate temperature range (50 K < $T$ < 200 K) is not associated with the current flow, but is due to the initial high magnetic ordering of stuck together ferromagnetic nanoparticles.
2. Depletion of the percolation cluster with decreasing temperature at $T = 10$ K causes nonlinearity of current-voltage characteristics and the appearance of a dependence of the magnitude of the coercivity fields of the conducting system of the sample on the measuring current $H_p(I)$, associated with the effects of overheating in the percolation cluster.
3. At the lowest temperatures ($T$ < 7 K), as a result of an increase in the degree of spin polarization ($P$), the slope of the dependencies for the fields of the coercive force $H_p(I)$ on the value of the flowing spin-polarized current $I$ turns out to be inverted, i.e., the coercivity field of the magnetoresistive hysteresis loops $H_p$ increases with increasing current. This effect is presumably due to the influence of the flowing spin-polarized current on the domain structure of large $CrO_2$ nanoparticles located in critical links connecting magnetically ordered percolation chains.

The authors of this paper acknowledge partial support from the NAS of Ukraine target program, "Prospective basic research and innovative development of nanomaterials and nanotechnologies for the needs of industry, healthcare, and agriculture" (Grant No. 2/22-H from the NAS of Ukraine).